\documentclass{iopart}
\usepackage{iopams}
\usepackage{graphicx}
\newcommand{\cc}[1]{\overline{#1}}
\newcommand{\cket}[1]{\vert #1 \rangle}
\newcommand{\bra}[1]{\langle #1 \vert}

\newcommand{\Id}{\mathbb{I}}

\begin{document}
\title{All degree six local unitary invariants of $k$ qudits}
\author{Szil\'ard Szalay}
\address{Department of Theoretical Physics,
Institute of Physics,
Budapest University of Technology and Economics,
H-1111 Budapest, Budafoki \'ut 8, Hungary}
\ead{szalay@phy.bme.hu}
\date{\today}
\begin{abstract}
We give explicit index-free formulae for
all the degree six (and also degree four and two) 
algebraically independent local unitary invariant polynomials for
finite dimensional $k$-partite pure and mixed quantum states.
We carry out this by the use of graph-technical methods,
which provides illustrations for this abstract topic.
\end{abstract}

\section{Introduction}
The notion of entanglement of a composite quantum system
is known to be invariant under unitary transformations on the subsystems,
so the investigation of local unitary (LU) invariants is a
natural way of studying quantum entanglement.
In this paper, we give illustrations for the general results of
Hero \textit{et. al.}~\cite{HW,HWW} and Vrana~\cite{Peti1,Peti23} 
on LU-invariant polynomials for pure quantum states.
In~\cite{Peti23}, 
it has been pointed out that the \emph{inverse limit} (in the local dimensions) of algebras of LU-invariant polynomials
of finite dimensional $k$-partite quantum systems is \emph{free,} 
and an \emph{algebraically independent generating set} for that has been given.
This approach using the inverse limit construction is different from the usual,
when the LU orbit structure is investigated first---for given local dimensions---and 
then invariants separating the orbits are being searched for 
\cite{RainsInv,LPpureorbs,LPSmixedorbs,GrasslLU,Sudbery3qb,Makhlin2qbMixed,Acin3qbPureCanon}.
The structure of algebras of LU-invariant polynomials for given local dimensions is very complicated,
the inverse limit of these~\cite{HW,HWW,Peti23}, however, has a remarkably simple structure: it is free~\cite{Peti23},
and an algebraically independent generating set can be given for that.
Moreover, from the results for \emph{pure} states, one can also obtain algebraically independent LU-invariant polynomials for \emph{mixed} states~\cite{Peti23}.

The aim of this paper is to draw the attention of researchers working in the field of quantum information 
to the approach above---provided by researchers with expertise on representation theory---by hinting at
the nature of results obtained within this approach.
In particular, 
we write out explicitly the \emph{linearly independent basis} of the inverse limit of algebras
and single out the members of the \emph{algebraically independent generating set} from them
in the first three graded subspace of the algebras. 
We give these polynomials in an index-free form for \emph{arbitrary number of subsystems.}

The outline of this paper is as follows.
In section~\ref{sec:luinvs}, we introduce the general writings of an LU-invariant polynomial
and preclude the appearance of identical ones in a less abstract way than was done originally in~\cite{HWW,Peti23}.
We discuss the cases of pure and mixed quantum states.
In section~\ref{sec:graphsops}, following~\cite{HWW,Peti23}, we introduce graphs for the LU-invariant polynomials.
Then we learn to read off matrix operations
(such as partial trace, matrix product, tensorial product or partial transpose)
from graphs.
If this can be done for a whole graph of an LU-invariant polynomial, 
then we can write a nice index-free formula for that by these operations.
In section~\ref{sec:pureinv}, we give these index-free formulae for pure state invariants of degree two, four and six.
Using graphs, these formulae can be given for arbitrary number of subsystems.
In section~\ref{sec:mixinv}, we show the formulae for mixed states
and we discuss the connection of pure and mixed quantum states from another point of wiev.
In section~\ref{sec:alg}, we give an algorithm for the construction of the 
labelling of different invariant polynomials of degree six.
(For degree two and four, this task is trivial.)
Summary and some notes are left for section~\ref{sec:summary}.

\section{Local unitary invariant polynomials}
\label{sec:luinvs}

\subsection{Invariants for pure states}
\label{ssec:luinvs:pure}

Let $\mathcal{H}=\mathcal{H}_1\otimes\dots\otimes\mathcal{H}_k$ be
the Hilbert-space of a $k$-partite composite system,
where $\dim \mathcal{H}_j=n_j$
and $n$ denotes the $k$-tuple of these local dimensions: $n=(n_1,\dots,n_k)$.
An element of the Hilbert-space can be written as
$\cket{\psi}=\psi_{i_1,\dots,i_k}\cket{i_1,\dots,i_k}$,
where $\cket{i_j}\in\mathcal{H}_j$ for $i_j=1,\dots,n_j$ is an orthonormal basis for all $1\leq j\leq k$,
and the summation over $i_j=1,\dots,n_j$ is understood.
As usual in the topic of quantum invariants, the norm of $\psi$ does not have to be fixed.

It is well-known (see e.g. in~\cite{Sudbery3qb}) that the way to get local unitary invariants is the following.
We write down the term $(\psi_{i_1,\dots,i_k}\cc{\psi}_{i'_1,\dots,i'_k})$ $m$ times
(with different indices)
and contract all primed indices with unprimed indices on the same $\mathcal{H}_j$.
A polynomial obtained in this way is of degree $2m$,%
---degree $m$ in the coefficients and also in their complex conjugates.
This is the only case in which unitary invariants can arise~\cite{Peti23},
so it is convenient to use this natural gradation,
and to call this polynomial of \emph{grade} $m$.
(In the case of mixed states the grade coincides with the degree in the matrix-elements of the density matrix.)
The possible index-contractions on an $\mathcal{H}_j$ are encoded by the elements of $S_m$,
the group of the permutations of $m$ letters.
$\sigma_j\in S_m$ tells us that the primed index of the $l$th term is contracted with the unprimed index of the $\sigma_j(l)$'th term,
so there is an index-contraction scheme for all $k$-tuples of permutations 
${\sigma}=(\sigma_1,\dots,\sigma_k) \in S_m^k$:
\begin{equation}
\label{purinv0}
f_{{\sigma}}(\psi)=
\psi_{i_1^1,\dots,i_k^1}\cdots
\psi_{i_1^m,\dots,i_k^m}
\cc{\psi}_{i_1^{\sigma_1(1)},\dots,i_k^{\sigma_{k}(1)}}\cdots
\cc{\psi}_{i_1^{\sigma_1(m)},\dots,i_k^{\sigma_{k}(m)}},
\end{equation}
where the summation
over $i_j^l=1,\dots,n_j$ for all $1\leq j\leq k$ and $1\leq l\leq m$ is understood.
(The lower labels of $i$'s refer to the subsystems
and the upper ones refer to the different index-contractions.)

However, different $k$-tuples of permutations can give rise to the same polynomial.
We have the terms $(\psi_{i_1^l,\dots}\cc{\psi}_{i_1^{\sigma_1(l)},\dots})$ $m$ times:
\begin{equation*}
 (\psi_{i_1^1,\dots}\cc{\psi}_{i_1^{\sigma_1(1)},\dots})
 (\psi_{i_1^2,\dots}\cc{\psi}_{i_1^{\sigma_1(2)},\dots})\dots
 (\psi_{i_1^m,\dots}\cc{\psi}_{i_1^{\sigma_1(m)},\dots}), 
\end{equation*}
but it makes no difference if we permute the $\psi_{i_1^l,\dots}$'s or $\cc{\psi}_{i_1^{\sigma_1(l)},\dots}$'s 
among these terms, since, being scalar variables, they commute.
This is equivalent to the relabelling of the indices (in the upper labels),
which can be formulated by the permutations $\alpha,\beta\in S_m$
encoding the permutations of $\cc{\psi}_{i_1^{\sigma_1(l)},\dots}$'s and $\psi_{i_1^l,\dots}$'s, respectively:
\begin{eqnarray*}
& (\psi_{i_1^1,\dots}\cc{\psi}_{i_1^{\sigma_1(1)},\dots})
  (\psi_{i_1^2,\dots}\cc{\psi}_{i_1^{\sigma_1(2)},\dots})\dots
  (\psi_{i_1^m,\dots}\cc{\psi}_{i_1^{\sigma_1(m)},\dots})\\
&=(\psi_{i_1^{\beta(1)},\dots}\cc{\psi}_{i_1^{\alpha\sigma_1(1)},\dots})
  (\psi_{i_1^{\beta(2)},\dots}\cc{\psi}_{i_1^{\alpha\sigma_1(2)},\dots})\dots
  (\psi_{i_1^{\beta(m)},\dots}\cc{\psi}_{i_1^{\alpha\sigma_1(m)},\dots})\\
&=(\psi_{i_1^1,\dots}\cc{\psi}_{i_1^{\alpha\sigma_1\beta^{-1}(1)},\dots})
  (\psi_{i_1^2,\dots}\cc{\psi}_{i_1^{\alpha\sigma_1\beta^{-1}(2)},\dots})\dots
  (\psi_{i_1^m,\dots}\cc{\psi}_{i_1^{\alpha\sigma_1\beta^{-1}(m)},\dots}).
\end{eqnarray*}
(Here we have written out only the indices on $\mathcal{H}_1$ to get shorter expressions,
but, obviously, the same $\alpha$ and $\beta$ work on every index running on every $\mathcal{H}_j$.)
Therefore we have
\begin{equation}
f_{(\sigma_1,\dots,\sigma_k)}(\psi)=
f_{(\alpha\sigma_1\beta^{-1},\dots,\alpha\sigma_k\beta^{-1})}(\psi),
\end{equation}
giving rise to an equivalence relation on $S_m^k$:
\begin{equation}
{\sigma}\sim{\sigma}' \;\;\mathrm{iff} \;\; \exists \alpha,\beta\in S_m: \sigma'_j=\alpha\sigma_j\beta^{-1}, 1\leq j\leq k
\end{equation}
and the equivalence classes are denoted by
$[{\sigma}]_\sim=[\sigma_1,\dots,\sigma_k]_\sim
=\{(\alpha\sigma_1\beta^{-1},\dots,\alpha\sigma_k\beta^{-1})\mid \alpha,\beta\in S_m\}$.
The set of these equivalence classes is the double-cosets of $S_m^k$ by the diagonal action:
$\Delta \backslash S_m^k /\Delta$,
 where the subgroup $\Delta=\{(\delta,\dots,\delta)\mid \delta\in S_m\}\subseteq S_m^k$.

Thus, the ambiguity arising from the commutativity of the $m$ terms $\psi_{\dots}$ and $\cc{\psi}_{\dots}$ 
in (\ref{purinv0}) has been handled
by the labelling of the polynomials by the elements of $\Delta \backslash S_m^k /\Delta$.
As a next step, it would be desirable to get one representing element for every equivalence class.
Unfortunately, this can not be done generally, (i.e., for an arbitrary $m$,)
but we can make the equivalence classes smaller by throwing off some of their elements in a general way.
Every equivalence class has elements having the identity permutation $e$ in the last position.
Indeed, we have $\alpha\sigma_k\beta^{-1}=e$ 
in $(\alpha\sigma_1\beta^{-1},\dots,\alpha\sigma_k\beta^{-1})$ if we set $\alpha=\beta\sigma_k^{-1}$:
\begin{eqnarray*}
(\alpha\sigma_1\beta^{-1},\dots,\alpha\sigma_{k-1}\beta^{-1},\alpha\sigma_k\beta^{-1})\sim\\
(\beta\sigma_k^{-1}\sigma_1\beta^{-1},\dots,\beta\sigma_k^{-1}\sigma_{k-1}\beta^{-1},e)=
(\beta\sigma'_1\beta^{-1},\dots,\beta\sigma'_{k-1}\beta^{-1},e),
\end{eqnarray*}
which is actually an orbit of $S_m^{k-1}\times\{e\}$ under the action of simultaneous conjugation.
So it is useful to define another equivalence relation on $S_m^k$
\begin{equation}
{\sigma}\approx{\sigma}' \;\;\mathrm{iff} \;\; \exists \beta\in S_m: \sigma'_j=\beta\sigma_j\beta^{-1}, 1\leq j\leq k
\end{equation}
and the equivalence classes are denoted by
$[{\sigma}]_\approx=[\sigma_1,\dots,\sigma_k]_\approx
=\{(\beta\sigma_1\beta^{-1},\dots,\beta\sigma_k\beta^{-1})\mid \beta\in S_m\}$.
The set of these equivalence classes is denoted by $S_m^k/S_m$.
This equivalence is defined on $S_m^{k-1}$ in the same way.
$S_m^{k-1}$ can be injected into $S_m^k$ by
$\imath:S_m^{k-1}\hookrightarrow S_m^k$ as $\imath(\sigma_1,\dots,\sigma_{k-1})=(\sigma_1,\dots,\sigma_{k-1},e)$,
which is compatible with the equivalence $\approx$, but not with $\sim$.
Since if ${\sigma}\approx{\sigma}'$ then ${\sigma}\sim{\sigma}'$,
a $\sim$-equivalence class is the union of disjoint $\approx$-equivalence classes
\begin{equation}
\label{classdecomp}
[{\sigma}]_\sim=[{\sigma}^{(1)}]_\approx\cup[{\sigma}^{(2)}]_\approx\cup\dots
\end{equation}
The elements of $[{\sigma}]_\sim$ which have $\sigma_k=e$ form one of the $\approx$-equivalence classes of the right-hand side.
This $\approx$-equivalence class (element of $S_m^{k-1}/S_m$) is also suitable for the labelling of the polynomials
instead of the original $\sim$--equivalence class (element of $\Delta \backslash S_m^k /\Delta$).

The meaning of the choice $\alpha\sigma_k\beta^{-1}=e$ is that
the indices on $\mathcal{H}_k$ are contracted \emph{inside}
every term $(\psi_{i_1^l,\dots,i_k^l}\cc{\psi}_{i_1^{\sigma_1(l)},\dots,i_k^l})$.
This ``couples together'' the pairs of $\psi$ and $\cc{\psi}$.
The simultaneous conjugation means the permutation of the $m$ terms $(\psi_{\dots}\cc{\psi}_{\dots})$,
which is the remaining ambiguity arising from the commutativity of these terms.
Note, that we have singled out the last Hilbert-space $\mathcal{H}_k$ in this construction.
In the general aspects, it makes no difference which Hilbert-space is singled out,
but as we write the pure-state invariants using matrix operations,
it can happen---and usually it will happen---that this freedom manifests itself
in the different writings of the same pure state invariant.

Summing up,
for a composite system of $k$ subsystems,
the LU-invariant polynomial given by $[\sigma_1,\dots,\sigma_{k-1}]_\approx \in S_m^{k-1}/S_m$ is
\begin{equation}
\label{purinv}
\fl
f_{[\sigma_1,\dots,\sigma_{k-1}]_\approx}(\psi)=
\psi_{i_1^1,\dots,i_k^1}\cdots
\psi_{i_1^m,\dots,i_k^m}
\cc{\psi}_{i_1^{\sigma_1(1)},\dots,i_{k-1}^{\sigma_{k-1}(1)},i_k^1}\cdots
\cc{\psi}_{i_1^{\sigma_1(m)},\dots,i_{k-1}^{\sigma_{k-1}(m)},i_k^m }.
\end{equation}
By the use of $S_m^{k-1}/S_m$ labelling,
we have got rid of the formal equivalence of polynomials arising from the commutativity of the terms,
and have got a set of LU-invariant polynomial for the elements of the set $S_m^{k-1}/S_m$.
Can it happen that different elements of $S_m^{k-1}/S_m$ gives the same polynomial?
Are there linear dependencies among these polynomials?
It is not known in general,
but sometimes there is more to be known:
(\ref{purinv}) gives a linearly independent basis in each $m$ graded subspace of the inverse limit of the algebras~\cite{HWW}.
Moreover,---as the main result of~\cite{Peti23} states,---%
an \emph{algebraically independent generating set} is formed by
the polynomials given in (\ref{purinv}) 
for which the defining $k-1$ permutations \emph{together} act transitively on the set of $m$ labels.
For the algebras of given local dimensions $n=(n_1,\dots,n_k)$, 
the above polynomials form a basis as long as $m\leq n_j$ (for all $j$),
otherwise they become linearly dependent.
The algebraic independency also fails if we restrict ourselves to given local dimensions.
(The algebra of LU-invariant polynomials is usually not even free for given local dimensions.)

\subsection{Invariants for mixed states}
\label{ssec:luinvs:mixed}

Now consider a mixed quantum state of the $k$-partite composite system.
This state is given by the density operator, acting on $\mathcal{H}$,
written as
$\varrho=\varrho_{i_1,\dots,i_k;i'_1,\dots,i'_k}\cket{i_1,\dots,i_k}\bra{i'_1,\dots,i'_k}$.
The density operator, by definition, a positive definite self adjoint operator,
but, as usual in the topic of quantum invariants, the trace of $\varrho$ does not have to be fixed.

The general form of an LU-invariant polynomial is given by a simillar index-contraction scheme,
 encoded by $\sigma=(\sigma_1,\dots,\sigma_k)\in S_m^k $, as in the case of pure states:
\begin{equation}
\label{mixinv0}
f_\sigma(\varrho)=
\varrho_{i_1^1,\dots,i_k^1;i_1^{\sigma_1(1)},\dots,i_k^{\sigma_{k}(1)}}\cdots
\varrho_{i_1^m,\dots,i_k^m;i_1^{\sigma_1(m)},\dots,i_k^{\sigma_{k}(m)}},
\end{equation}
where the summation
over $i_j^l=1,\dots,n_j$ for all $1\leq j\leq k$ and $1\leq l\leq m$ is understood.
(We denote the pure and the mixed state invariants with the same symbol,
the distinction between them is their arguments: they are vectors and matrices, respectively.)

Here we can carry out a similar construction as in the case of pure states,
with one difference:
the building blocks of the polynomials 
are the $(\varrho_{i_1,\dots, i_k;i'_1,\dots,i'_k})$ matrix-elements of the density operator
instead of the former $(\psi_{i_1,\dots,i_k}\cc{\psi}_{i'_1,\dots,i'_k})$'s.
Hence there is no step corresponding to the ``double coset'' construction:
we can not move the ``two parts'' of $\varrho$ independently as has been done in the case of $\psi\cc{\psi}$,
since in general $\varrho$ is not of rank one.
This means that we can not relabel the primed and unprimed indices independently.
The possible relabelling is given by $\beta\in S_m$:
\begin{eqnarray*}
& (\varrho_{i_1^1,\dots;i_1^{\sigma_1(1)},\dots})
  (\varrho_{i_1^2,\dots;i_1^{\sigma_1(2)},\dots})\dots
  (\varrho_{i_1^m,\dots;i_1^{\sigma_1(m)},\dots})\\
&=(\varrho_{i_1^{\beta(1)},\dots;i_1^{\beta\sigma_1(1)},\dots})
  (\varrho_{i_1^{\beta(2)},\dots;i_1^{\beta\sigma_1(2)},\dots})\dots
  (\varrho_{i_1^{\beta(m)},\dots;i_1^{\beta\sigma_1(m)},\dots})\\
&=(\varrho_{i_1^1,\dots;i_1^{\beta\sigma_1\beta^{-1}(1)},\dots})
  (\varrho_{i_1^2,\dots;i_1^{\beta\sigma_1\beta^{-1}(2)},\dots})\dots
  (\varrho_{i_1^m,\dots;i_1^{\beta\sigma_1\beta^{-1}(m)},\dots}).
\end{eqnarray*}
Therefore we have
\begin{equation}
f_{(\sigma_1,\dots,\sigma_k)}(\varrho)=
f_{(\beta\sigma_1\beta^{-1},\dots,\beta\sigma_k\beta^{-1})}(\varrho),
\end{equation}
the elements of the orbits in $S_m^k$ under the action of simultaneous conjugation
gives the same polynomial.
Let these orbits be denoted by
$[\sigma_1,\dots,\sigma_k]_\approx$,
as before, and the LU-invariant polynomial given by this is
\begin{equation}
\label{mixinv}
f_{[\sigma_1,\dots,\sigma_k]_\approx}(\varrho)=
\varrho_{i_1^1,\dots,i_k^1;i_1^{\sigma_1(1)},\dots,i_k^{\sigma_k(1)}}\cdots
\varrho_{i_1^m,\dots,i_k^m;i_1^{\sigma_1(m)},\dots,i_k^{\sigma_k(m)}}.
\end{equation}

The independency of these follows from the independency of the pure state invariants
when that is the case for the latter ones.
This is because we can obtain the independent mixed state invariants
of the system with local dimensions $n=(n_1,\dots,n_k)$,
if we add a large enough $\mathcal{H}_{k+1}$ Hilbert-space,
and calculate the invariants (\ref{purinv}) for a pure state $\cket{\phi}\in \mathcal{H}\otimes\mathcal{H}_{k+1}$.
(See \cite{Peti23} for the abstract construction.)
Since in (\ref{purinv}) we have not permuted the last (this time $k+1$'th) indices, 
we can read off the invarians for $\varrho=\Tr_{k+1}\cket{\phi}\bra{\phi}$ from (\ref{purinv}).
(If $\dim\mathcal{H}_{k+1}\geq\prod_{j=1}^k\dim\mathcal{H}_j$, then $\varrho$ can be of full rank,
and we can get all $\varrho$ acting on $\mathcal{H}$ in this way.)
Note that if we simply substitute $\varrho$ by a pure state $\cket{\psi}\bra{\psi}$ in (\ref{mixinv}),
then we do not get a linearly independent set of $k$-partite pure state invariants
for all the labels $[\sigma_1,\dots,\sigma_k]_\approx\in S_m^k/S_m$.
However, if we restrict this for the case when $\sigma_k=e$, then we get back the
linearly independent set of pure state invariants from the linearly independent set of mixed state ones of a $k$-partite system:
\begin{equation}
\label{purmix}
\fl
 f_{[\sigma_1,\dots,\sigma_{k-1}]_\approx}(\psi)
=      f_{[\sigma_1,\dots,\sigma_{k-1},e]_\approx}(      \cket{\psi}\bra{\psi})
\equiv f_{[\sigma_1,\dots,\sigma_{k-1}  ]_\approx}(\Tr_k \cket{\psi}\bra{\psi}).
\end{equation}

\section{Graphs and matrix operations}
\label{sec:graphsops}

\subsection{Graphs of invariants}
The index-contraction scheme of the LU-invariant polynomials given in the previous section
can be made more expressive by the use of graphs~\cite{HWW,Peti23}.
For a grade $m$ invariant, given by $\sigma=(\sigma_1,\dots,\sigma_k)\in S_m^k$,
one can draw a graph with $m$ vertices 
with the labels $1\leq l\leq m$.
These vertices represent 
the $m$ terms $(\psi_{i_1^l,\dots,i_k^l}\cc{\psi}_{i_1^{\sigma_1(l)},\dots,i_k^{\sigma_k(l)}})$
in (\ref{purinv0}).
The edges of the graph are directed and coloured with $k$ different colours.
The edges of the $j$'th colour encode
the index contractions on the $j$'th Hilbert-space
given by the permutation $\sigma_j$ of $(\sigma_1,\dots,\sigma_k)$:
for every $1\leq l\leq m$ there is an edge with head and tail on the $l$'th and $\sigma_j(l)$'th vertex, respectively,
meaning a contracted $j$'th index of the $l$'th $\psi$ and the $\sigma_j(l)$'th $\cc{\psi}$.

As we have seen in the previous section,
some elements of $S_m^k$ give rise to the same polynomials, as was elaborated in the previous section.
How can we tell that story in the language of graphs?
Following the previous section, for a given $\sigma\in S_m^k$,
the elements of the $\sim$-equivalence class $[\sigma]_\sim \in \Delta\backslash S_m^k/\Delta$ give the same invariant.
First we set $\sigma_k=e$,
which means that we select the graphs having a loop of colour $k$ on every vertex.
For a given $\sim$-equivalence class, there are still many graphs of that kind,
and they are given by the elements of the corresponding $\approx$-equivalence class.
How are these graphs related to each other?
The simultaneously conjugation by a $\beta\in S_m$ means the relabelling of the vertices,
that is, the relabelling of the indices (in the upper label) 
of the terms $(\psi_{i_1^l,\dots,i_k^l}\cc{\psi}_{i_1^{\sigma_1(l)},\dots,i_k^{\sigma_k(l)}})$.
So the elements of a $\approx$-equivalence class give the same graph with all the possible labellings,
and the $\approx$-equivalence class itself gives an \emph{unlabelled} graph.

Since the elements of $S_m^k$ related by simultaneous conjugation gives rise to the same unlabelled graph,
the decomposition in (\ref{classdecomp}) shows that there may exist
many unlabelled graphs (many $\approx$-classes) giving rise to the same polynomial defined by a given $\sim$-class.
For example, 
we can set $\sigma_j=e$ for a $j\neq k$,
which results graphs where the edges of colour $j\neq k$ form loops on every vertex.
On the other hand, there may be $\approx$-classes in the given $\sim$-class which does not contain $e$.
All of these graphs give the same polynomial, but it can happen that
some of them can be formulated using matrix operations (in different ways for different graphs)
and some of them not.
(It turns out (see in next section) that every polynomial can be formulated using matrix operations up to $m=3$.)

The case of mixed states is simpler, because there are no $\sim$-classes involved.
For a grade $m$ invariant given by $\sigma\in S_m^k$,
the vertices represent the terms $(\varrho_{i_1^l,\dots,i_k^l;i_1^{\sigma_1(l)},\dots,i_k^{\sigma_k(l)}})$,
and only the polynomials given by the elements of the $\approx$-equivalence class $[\sigma]_\approx\in S_m^k/S_m$
are the same by the commutativity of these terms.
This means that we simply throw off the labelling of the vertices of the graph given by $\sigma$.

\subsection{Graphs of matrix operations}

The building blocks
$(\psi_{i_1,\dots,i_k}\cc{\psi}_{i'_1,\dots,i'_k})$ and $(\varrho_{i_1,\dots,i_k;i'_1,\dots,i'_k})$
of the polynomials
are matrices 
with row and column indices being the unprimed and primed ones, respectively.
So we expect that some of the invariant polynomials can be written using only matrix operations,
such as partial trace, matrix products, tensorial products or partial transpose.
How can we read off matrix operations from the graphs corresponding to the invariant polynomials?
This is a difficult question in general and---as we will see---not all graphs can be encoded using matrix operations.
It is more instructive to look at the graphs coming from the matrix operations first,
and then to search for these elementary subgraphs in a general graph coming from a polynomial given by an element of $S_m^k$.

\begin{figure}[ht]
\includegraphics{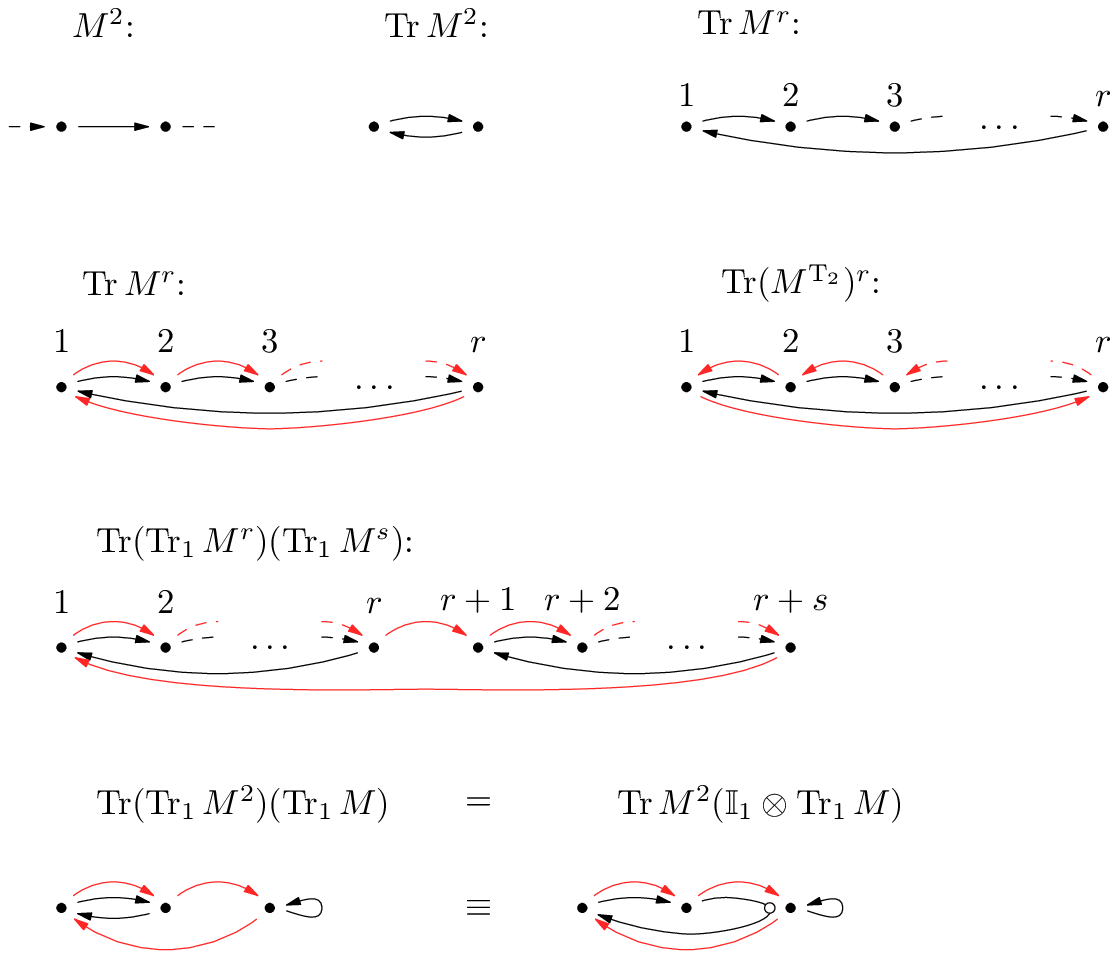}\centering
\caption{Elementary matrix operations represented by graphs.
In the first row: $k=1$, there is only one colour of edges representing the index contractions.
In the other rows: $k=2$, two different colours of edges correspond to the index contractions on the two Hilbert-spaces:
black and red on $\mathcal{H}_1$ and $\mathcal{H}_2$, respectively.
(Black and dark grey in greyscale printing.)
 }\label{fig:mxopgraphs}
\end{figure}

Let us see some matrix operations and their graphs.
The matrix multiplication means 
contraction of the column indices of the first matrix with the row indices of the second matrix,
the trace means contraction of the column indices with the row indices
and the partial transposition means the swap of the given row and column indices.
First consider only the indices belonging to the Hilbert space of only the first subsystem---i.e., we have edges of only one colour.
For a general matrix $M$, which is represented by the vertices of the graphs, 
the multiplicaton by itself gives the edge from one vertex to another,
the $r$'th power $M^r$ is a chain of edges (without loops),
and the trace of it closes this chain into a loop. (See in the first row of figure~\ref{fig:mxopgraphs}.)
Now let us take into account indices belonging to the second subsystem.
$\Tr M^r$ is the same loop as before but with doubled edges,
while the partial transposition $\Tr(M^{T_2})^r$ reverses the loop of the corresponding colour
(second row of figure~\ref{fig:mxopgraphs}).
The partial traces in $\Tr \Tr_1 (M^r) \Tr_1 (M^s)$  make smaller loops on a subsystem
(third row of figure~\ref{fig:mxopgraphs}).
There is a little trick, which is prooved to be very useful later: 
$\Tr (\Tr_1 M^2) (\Tr_1 M)=\Tr M^2(\Id_1\otimes \Tr_1 M)$.
On the language of graphs we just bend the corresponding edge next to the vertex representing $\Tr_1 M$,
and we draw a circle on it, representing the identity matrix, which is just contract indices
(last row of figure~\ref{fig:mxopgraphs}).
If the graph is the union of disjoint graphs, then the corresponding polynomial is factorizable, 
since the summations corresponding to the disjoint pieces can be carried out independently.
This almost trivial situation is getting more complicated,
if we take into account the indices of all the subsystems---i.e., the edges of all colours.
Examples are shown in the next section.

\section{Pure state invariants}
\label{sec:pureinv}

In the following, we illustrate how a pure state LU-invariant polynomial 
(encoded by $[{\sigma}]_\sim\in\Delta \backslash S_m^k /\Delta$)
is given by different unlabelled graphs (encoded by $[{\sigma}]_\approx\in S_m^k/S_m$).
(While an unlabelled graph 
is given by different labelled graphs (index-contraction scheme, encoded by ${\sigma}\in S_m^k$).)
On the other hand,
different unlabelled graphs give rise to different writings by matrix operations of the same polynomial.
The polynomials are labelled here by the elements of $S_m^{k-1}/S_m$ 
instead of the elements of $\Delta \backslash S_m^k /\Delta$,
which give special unlabelled graphs having loops of colour $k$ on every vertex.
(Sometimes, e.g., in \cite{HWW}, these loops are omitted, and only the first $k-1$-coloured edges are drawn.)
For a permutation $k-1$-tuple ${\sigma}\in S_m^{k-1}$,
$[\sigma]_\approx\in S_m^{k-1}/S_m$,
and we can write for the corresponding invariant
$[\imath({\sigma})]_\sim
=[\imath({\sigma})]_\approx\cup
[{\sigma}^{(2)}]_\approx\cup
[{\sigma}^{(3)}]_\approx\cup\dots$,
where ${\sigma}^{(2)},{\sigma}^{(3)},\dots\in S_m^k$ are
representing elements of $\approx$-classes giving different graphs for the same invariant.

Let us see how these technics work.
As a warm-up, we show for all $k$ the trivial case of $m=1$ and the almost trivial case of $m=2$.
This is followed by the case of $m=3$, which is more interesting
because of the non-Abelian structure of $S_3$.
This is done for all $k$ too.
Let $\cket{\psi}\in\mathcal{H}$,
and, as we have seen, everything can be formulated using the
rank-one projector $\pi\equiv\pi_{12\dots k}=\cket{\psi}\bra{\psi}$.
We denote the reduced density matrices with the label of subsystems
which are not traced out, 
for example $\pi_{2\dots k}=\Tr_1 \pi_{12\dots k}$ etc.

\subsection{Invariant polynomials of grade $m=1$ (degree two)}
\begin{figure}[ht]
\includegraphics{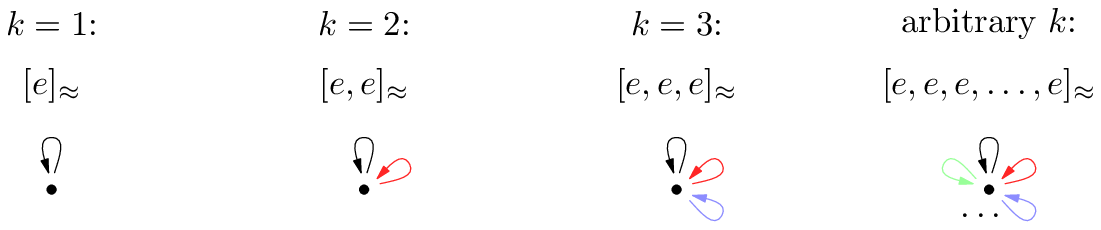}\centering
\caption{Graphs corresponding to the $m=1$ invariant polynomials.
 Black, red, blue and green edges (from the darkest to the lightest in greyscale printing)
 represent index-contractions on the first, second, third and last Hilbert spaces, respectively.}\label{fig:m1k}
\end{figure}

For $m=1$, we have the trivial $S_1=\{e\}$,
and for all $k$ number of subsystems $[e,e,\dots,e]_\sim=[e,e,\dots,e]_\approx$
($\Delta \backslash S_1^k /\Delta\cong S_1^k/S_1\cong S_1$)
meaning only one kind of graphs, having only one vertex.
Every edge---of $k$ different colors for the $k$ subsystems---starts and ends here.
(See in figure~\ref{fig:m1k}.)
This graphs mean a simple trace, which is the only possible index contraction.
The label of the polynomial is the only one element of $S_1^{k-1}/S_1\cong S_1$, and
\begin{equation}
\label{purinv1k}
f_{[e,\dots,e]_\approx}(\psi) =  \Tr\pi_{12\dots k} =\Vert\psi\Vert^2.
\end{equation}

\subsection{Invariant polynomials of grade $m=2$ (degree four)}
For $m=2$, we have $S_2=\{e,t\}$
(where $e=(1)(2)$ and $t=(12)$)
with the conjugacy-classes $[e]$ and $[t]$,
so the labels of the polynomials are $S_2^{k-1}/S_2\cong S_2^{k-1}$ for all $k$.
On the other hand, $[{\sigma}]_\sim=[{\sigma}]_\approx\cup [\overline{{\sigma}}]_\approx $,
(where $\overline{{\sigma}}_i= \overline{ {\sigma}_i}$, and $\overline{t}=e$, $\overline{e}=t$)
so there are two kinds of graphs for every polynomial.

\begin{figure}[ht]
\includegraphics{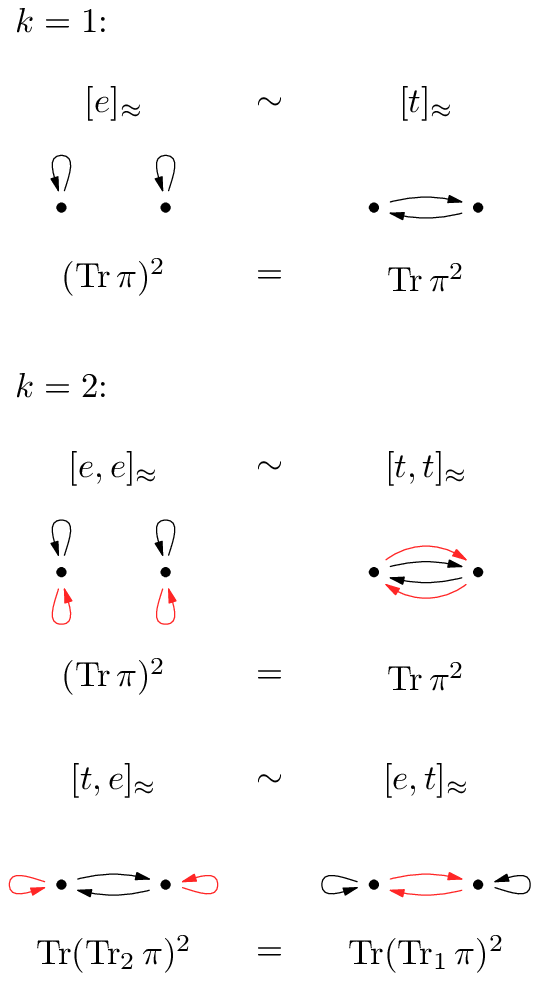}\centering
\caption{Graphs corresponding to the $m=2$ invariant polynomials for $k=1$ and $2$.
 Black and red edges (black and dark grey in greyscale printing)
 represent index-contractions on the first and second Hilbert spaces, respectively.}\label{fig:m2k12}
\end{figure}

For one-partite system, ($k=1$, $\pi\equiv\pi_1$,) 
the only polynomial is for
\begin{equation*}
 {}[e]_\sim=[e]_\approx\cup[t]_\approx.
\end{equation*}
From its graphs, (see in figure~\ref{fig:m2k12},) we have
\begin{equation*}
 f_{[]_\approx}(\psi) = (\Tr\pi_1)^2 = \Tr\pi_1^2 = \Vert\psi\Vert^4.
\end{equation*}

For two-partite system, ($k=2$, $\pi\equiv\pi_{12}$,)
there are two linearly independent polynomials. These are given by
\begin{eqnarray*}
 {}[e,e]_\sim &=[e,e]_\approx\cup[t,t]_\approx,\\
 {}[t,e]_\sim &=[t,e]_\approx\cup[e,t]_\approx.
\end{eqnarray*}
From their graphs, (see in figure~\ref{fig:m2k12},) we have
\begin{eqnarray*}
 f_{[e]_\approx}(\psi) &= (\Tr\pi_{12})^2 = \Tr\pi_{12}^2 = \Vert\psi\Vert^4,\\
 f_{[t]_\approx}(\psi) &= \Tr\pi_1^2 = \Tr\pi_2^2.
\end{eqnarray*}

For three-partite system, ($k=3$, $\pi\equiv\pi_{123}$,)
there are four linearly independent polynomials. These are given by
\begin{eqnarray*}
 {}[e,e,e]_\sim &=[e,e,e]_\approx\cup[t,t,t]_\approx,\\
 {}[e,t,e]_\sim &=[e,t,e]_\approx\cup[t,e,t]_\approx,\\
 {}[t,e,e]_\sim &=[t,e,e]_\approx\cup[e,t,t]_\approx,\\
 {}[t,t,e]_\sim &=[t,t,e]_\approx\cup[e,e,t]_\approx.
\end{eqnarray*}
From their graphs, we have
\begin{eqnarray*}
 f_{[e,e]_\approx}(\psi) &= (\Tr\pi_{123})^2 = \Tr\pi_{123}^2 = \Vert\psi\Vert^4,\\
 f_{[e,t]_\approx}(\psi) &= \Tr\pi_2^2 = \Tr\pi_{13}^2,\\
 f_{[t,e]_\approx}(\psi) &= \Tr\pi_1^2 = \Tr\pi_{23}^2,\\
 f_{[t,t]_\approx}(\psi) &= \Tr\pi_{12}^2 = \Tr\pi_3^2.
\end{eqnarray*}

The construction of these formulae can be easily generalized to arbitrary number of subsystems.
For this, take a look at the left graph of the last line of figure~\ref{fig:m2k12}.
This time, let the 
red lines (dark grey in black and white printing) 
represent the index-contractions on \emph{all} Hilbert-spaces
on which $\sigma_j=e$,
and the black lines
represent the index-contractions on \emph{all} Hilbert-spaces
on which $\sigma_j=t$.
Thus, we can read off the matrix operations for arbitrary $k$.
The other way of writing the polynomial can be reached by the interchange of the roles of the
black and red lines.
So, for arbitrary number of subsystems ($\pi\equiv\pi_{12\dots k}$),
for the polynomials for $[{\sigma}]_\sim=[{\sigma}]_\approx\cup [\overline{{\sigma}}]_\approx$,
we have
\begin{equation}
\label{purinv2k}
 f_{[\sigma_1,\dots,\sigma_{k-1}]_\approx}(\psi)
= \Tr(\Tr_{\{k\}\cup\{j\mid \sigma_j=e\}}\pi)^2
= \Tr(\Tr_{\{  j\mid \sigma_j=t\}}\pi)^2.
\end{equation}
This was well-known for qubits~\cite{Sudbery3qb}.
The number of these---the dimension of the grade $m=2$ subspace of the inverse limit of the algebras---is $2^{k-1}$.
The set of algebraically independent generators
contains all the $m=2$ polynomials from (\ref{purinv2k}), 
except the ones for which there are only $e$'s in $[\sigma_1,\dots,\sigma_{k-1}]_\approx$ labelling the polynomial.
(This is the only way for the permutations not to act transitively on the set of $m=2$ labels.)
The number of these is $2^{k-1}-1$.

\subsection{Invariant polynomials of grade $m=3$ (degree six)}
For $m=3$, we have $S_3=\{e,s,s^2,t,ts,ts^2\}$
(where $e=(1)(2)(3)$, $t=(12)(3)$, and $s=(123)$)
with the conjugacy-classes $[e]=\{e\}$, $[s]=\{s,s^2\}$ and $[t]=\{t,ts,ts^2\}$.
This time, we have no simple general rule for the splitting of a $\sim$-class to $\approx$-classes.

\begin{figure}
\includegraphics{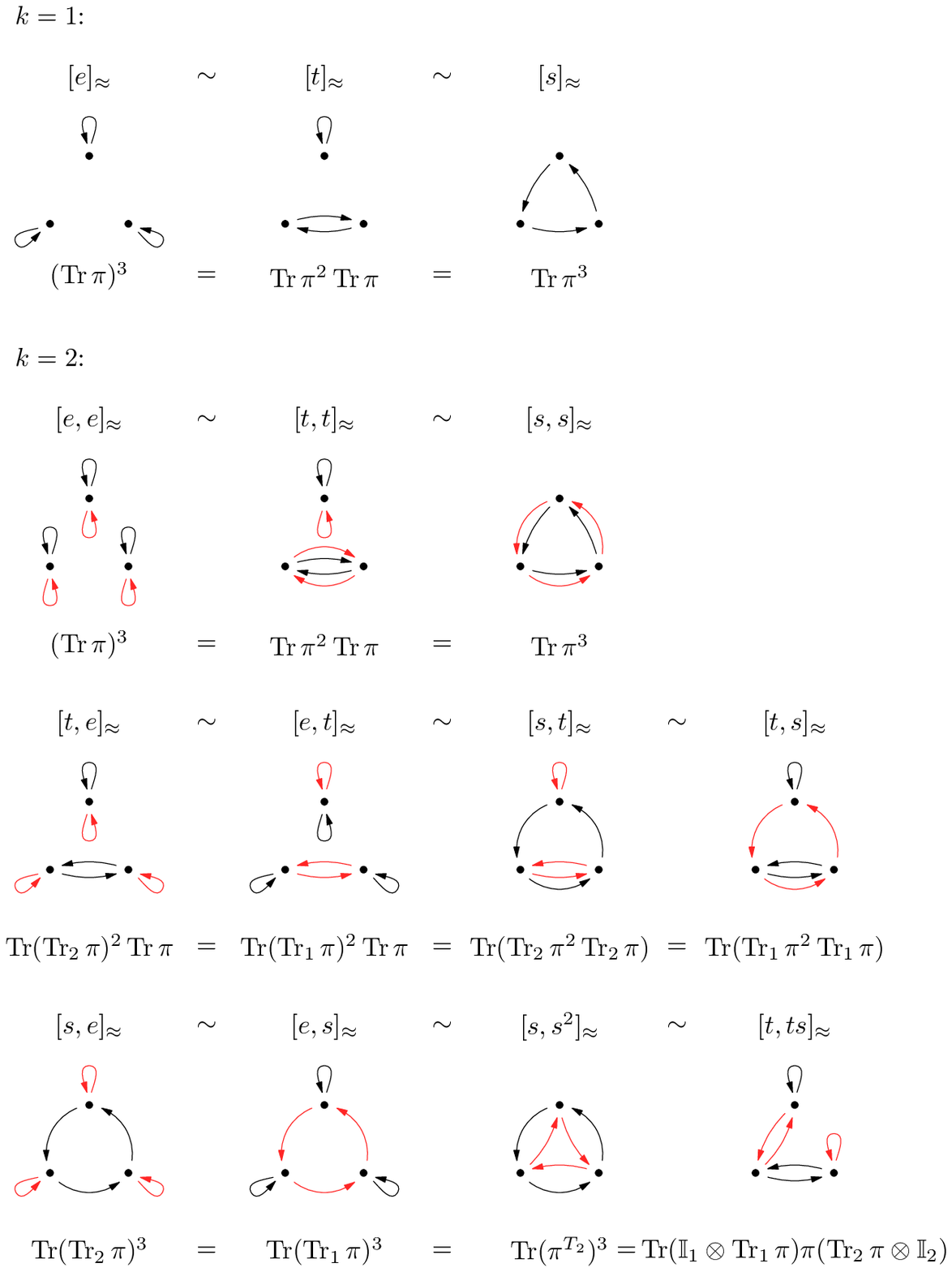}\centering
\caption{ Graphs corresponding to the $m=3$ invariant polynomials for $k=1$ and $2$.
 Black and red edges (black and dark grey in greyscale printing)
 represent index-contractions on the first and second Hilbert spaces, respectively.
 The formulae of the polynomials given by matrix operations, which can be red off from the graphs, are also written out.
 For the last one, we have use the trick in the last line of figure~\ref{fig:mxopgraphs} twice.}\label{fig:m3k12}
\end{figure}

For one-partite system, ($k=1$, $\pi\equiv\pi_1$,) 
the only polynomial is for
\begin{equation*}
 {}[e]_\sim=[e]_\approx\cup[t]_\approx\cup[s]_\approx.
\end{equation*}
From its graphs, (see in figure~\ref{fig:m3k12},) we have
\begin{equation*}
 f_{[]_\approx}(\psi) = (\Tr\pi_1)^3 = \Tr\pi_1^2\Tr\pi_1 = \Tr\pi_1^3 = \Vert\psi\Vert^6.
\end{equation*}

For two-partite system, ($k=2$, $\pi\equiv\pi_{12}$,) 
there are three linearly independent polynomials. These are given by
\begin{eqnarray*}
 {}[e,e]_\sim &=[e,e]_\approx\cup[t,t]_\approx\cup[s,s]_\approx,\\
 {}[t,e]_\sim &=[t,e]_\approx\cup[e,t]_\approx\cup[s,t]_\approx\cup[t,s]_\approx,\\
 {}[s,e]_\sim &=[s,e]_\approx\cup[e,s]_\approx\cup[s,s^2]_\approx\cup[t,ts]_\approx
\end{eqnarray*}
(so $S_3^1/S_3\simeq \{[e],[s],[t]\}$).
From their graphs, (see in figure~\ref{fig:m3k12},) we have
\begin{eqnarray*}
 f_{[e]_\approx}(\psi) &= (\Tr\pi_{12})^3 = \Tr\pi_{12}^2\Tr\pi_{12} = \Tr\pi_{12}^3 = \Vert\psi\Vert^6,\\
 f_{[t]_\approx}(\psi) &= (\Tr\pi_a)^2\Tr\pi_{12} = \Tr(\Tr_a\pi_{12}^2) \pi_b, \\
 f_{[s]_\approx}(\psi) &= \Tr\pi_a^3 = \Tr(\pi_{12}^{T_a})^3 = \Tr(\pi_1\otimes\pi_2)\pi_{12},
\end{eqnarray*}
for all $a,b\in \{1,2\}$, $a\neq b$.

For three-partite system, ($k=3$, $\pi\equiv\pi_{123}$,) 
it turns out that there are eleven linearly independent polynomials. These are given by
\begin{eqnarray*}
 {}[e,e,e]_\sim &=[e,e,e]_\approx\cup[t,t,t]_\approx\cup[s,s,s]_\approx,\\
 {}[e,t,e]_\sim &=[e,t,e]_\approx\cup[t,e,t]_\approx\cup[s,t,s]_\approx\cup[t,s,t]_\approx,\\
 {}[t,e,e]_\sim &=[t,e,e]_\approx\cup[e,t,t]_\approx\cup[t,s,s]_\approx\cup[s,t,t]_\approx,\\
 {}[t,t,e]_\sim &=[t,t,e]_\approx\cup[e,e,t]_\approx\cup[s,s,t]_\approx\cup[t,t,s]_\approx,\\
 {}[e,s,e]_\sim &=[e,s,e]_\approx\cup[s,e,s]_\approx\cup[s,s^2,s]_\approx\cup[t,ts,t]_\approx,\\
 {}[s,e,e]_\sim &=[s,e,e]_\approx\cup[e,s,s]_\approx\cup[s^2,s,s]_\approx\cup[ts,t,t]_\approx,\\
 {}[s,s,e]_\sim &=[s,s,e]_\approx\cup[e,e,s]_\approx\cup[s,s,s^2]_\approx\cup[t,t,ts]_\approx,\\
 {}[s,s^2,e]_\sim &=[s,s^2,e]_\approx\cup[s,e,s^2]_\approx\cup[e,s,s^2]_\approx\cup[t,ts,ts^2]_\approx,\\
 {}[t,s,e]_\sim &=[t,s,e]_\approx\cup[t,e,s]_\approx\cup[e,t,ts]_\approx\cup[t,s,s^2]_\approx\cup[s,t,ts]_\approx\cup[s,t,ts^2]_\approx,\\
 {}[s,t,e]_\sim &=[s,t,e]_\approx\cup[e,t,s]_\approx\cup[t,e,ts]_\approx\cup[s,t,s^2]_\approx\cup[t,s,ts]_\approx\cup[t,s,ts^2]_\approx,\\
 {}[t,ts,e]_\sim &=[t,ts,e]_\approx\cup[e,s,t]_\approx\cup[s,e,t]_\approx\cup[s,s^2,t]_\approx\cup[t,ts,s]_\approx\cup[t,ts^2,s]_\approx.
\end{eqnarray*}
Here we do not write out all the 49 formulae for the graphs coming from the $\approx$-classes above,
we just show two or three of them for every polynomial.
\begin{eqnarray*}
 f_{[e,e]_\approx}(\psi) &= \Vert\psi\Vert^6 = (\Tr\pi_{123})^3,\\
 f_{[e,t]_\approx}(\psi) &=  \Tr\pi_{123} \Tr\pi_2^2 \equiv \Tr\pi_{123} \Tr\pi_{13}^2,\\
 f_{[t,e]_\approx}(\psi) &=  \Tr\pi_{123} \Tr\pi_1^2 \equiv \Tr\pi_{123} \Tr\pi_{23}^2,\\
 f_{[t,t]_\approx}(\psi) &=  \Tr\pi_{123} \Tr\pi_{12}^2 \equiv \Tr\pi_{123} \Tr\pi_3^2,\\
 f_{[e,s]_\approx}(\psi) &= \Tr\pi_2^3 \equiv \Tr\pi_{13}^3,\\
 f_{[s,e]_\approx}(\psi) &= \Tr\pi_1^3 \equiv \Tr\pi_{23}^3,\\
 f_{[s,s]_\approx}(\psi) &= \Tr\pi_{12}^3 \equiv \Tr\pi_3^3,\\
 f_{[t,s]_\approx}(\psi) &= \Tr(\Id_1\otimes\pi_2)\pi_{12}^2 \equiv \Tr(\Id_1\otimes\pi_3)\pi_{13}^2
                 \equiv \Tr(\pi_2\otimes\pi_3)\pi_{23}, \\
 f_{[s,t]_\approx}(\psi) &= \Tr(\pi_1\otimes\Id_2)\pi_{12}^2 \equiv \Tr(\Id_2\otimes\pi_3)\pi_{23}^2
                 \equiv  \Tr(\pi_1\otimes\pi_3)\pi_{13},\\
 f_{[t,ts]_\approx}(\psi) &=\Tr(\pi_1\otimes\pi_2)\pi_{12}
                  \equiv \Tr(\pi_1\otimes\Id_3)\pi_{13}^2 \equiv \Tr(\pi_2\otimes\Id_3)\pi_{23}^2,\\
 f_{[s,s^2]_\approx}(\psi) &= \Tr(\pi_{ab}^{T_a})^3 \quad a,b\in\{1,2,3\}\; {\rm distinct}.
\end{eqnarray*}
The last one of these is the Kempe-invariant~\cite{Kempe3qb},
which has arisen in the context of hidden nonlocality.
Kempe has defined this for $n=(2,2,2)$, i.e.,~three qubits,
but it can be written by her definition (see (17) in~\cite{Kempe3qb}) for all $n=(n_1,n_2,n_3)$ three qudit systems.
It is observed by Sudbery~\cite{Sudbery3qb} that for three qubits the Kempe-invariant
can be expressed as
$f_{[s,s^2]_\approx}(\psi) 
=3f_{[t,s]_\approx}(\psi)-f_{[e,s]_\approx}(\psi)-f_{[s,s]_\approx}(\psi)
=3f_{[s,t]_\approx}(\psi)-f_{[s,e]_\approx}(\psi)-f_{[s,s]_\approx}(\psi)
=3f_{[t,ts]_\approx}(\psi)-f_{[e,s]_\approx}(\psi)-f_{[s,e]_\approx}(\psi)$.
However, this is only for qubits: 
if $m\leq n_j$ for all $j$ (so at least qutrits)
then the 11 polynomials listed above are linearly independent.
Another important three-qubit permutation- and LU-invariant polynomial of degree six,
which has arisen in twistor-geometric~\cite{Peter3qbGeom} 
and Freudenthal~\cite{DuffFreudenthal3qb} approach of three-qubit entanglement,
is the norm square of the Freudenthal-dual of $\psi$.
It can be written as
$6\Vert T(\psi,\psi,\psi)\Vert^2
=4f_{[s,s^2]_\approx}(\psi) +5f_{[e,e]_\approx}(\psi) 
- 3f_{[e,t]_\approx}(\psi)-3f_{[t,e]_\approx}(\psi)-3f_{[t,t]_\approx}(\psi)$.
Note, that this expression is not unique, 
since these $f_{[\sigma_1,\sigma_2]_\approx}$ polynomials are not linearly independent in the case of three qubits.
 
\begin{figure}
\includegraphics{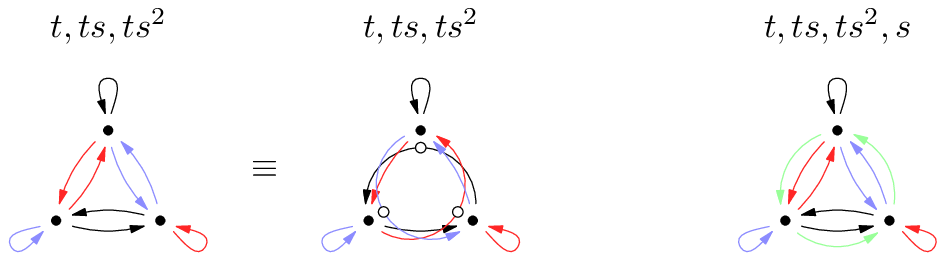}\centering
\caption{Graphs corresponding to the $m=3$ invariant polynomials.
 Black, red, blue and green edges (from the darkest to the lightest in greyscale printing)
 represent index-contractions on the Hilbert spaces on which $\sigma_j=t$, $ts$, $ts^2$ and $s$ respectively.
 For the first graph, we show how the trick in the last line of figure~\ref{fig:mxopgraphs} was used three times.}\label{fig:m3k}
\end{figure}

It is not obvious, but the construction of the grade $m=3$ polynomials
can be generalized to arbitrary number of subsystems.
To do this, consider an invariant given by ${\sigma}\in S_3^k$, where $\sigma_j\in\{t,ts,ts^2\}$.
This can be seen in figure~\ref{fig:m3k},
with the evident redefinitions of the meaning of the colours:
let the
black, red, and blue edges (black, dark grey and light grey in greyscale printing)
represent the index-contractions on \emph{all} Hilbert-spaces
on which $\sigma_j=t$, $ts$ and $ts^2$, respectively.
Using the trick in the last line of figure~\ref{fig:mxopgraphs} three times, we have, that
the polynomial is given by
the trace of the product of the three factors:
$\Id_{\{j\mid\sigma_j=t\}}\otimes\Tr_{\{j\mid\sigma_j=t\}}\pi$,
and other two with $ts$ and $ts^2$.
Note, that the order of these are arbitrary,
since it is related to the relabelling of the vertices of the graph.
However, if there are some subsystems on which $\sigma_j=s$,
that fixes the order of these terms up to cyclic permutation.
It turns out that we have to use the reverse ordering in the product,
the terms $\Id_{\{j\mid\sigma_j=\tau\}}\otimes\Tr_{\{j\mid\sigma_j=\tau\}}\pi$
with $\tau=ts^2$ first, then with $\tau=ts$ and then with $\tau=t$,
since they have the fixed point $1$, $2$ and $3$, respectively.
On the subsystems on which $\sigma_j=s$,
the indices are intact,---partial traces act only on other subsystems,---they are contracted in the appropriate way. 
If there are some subsystems on which $\sigma_j=s^2$,
then we use $\pi^{T_{\{j\mid\sigma_j=s^2\}}}$ instead of $\pi$ to reduce the situation to the known case.
Similarly, if there are some subsystems on which $\sigma_j=e$,
then we use $\Tr_{\{j\mid\sigma_j=e\}}\pi$ instead of $\pi$.
Summing up, 
for arbitrary number of subsystems ($\pi\equiv\pi_{12\dots k}$)
we have the following formula for the $m=3$ polynomials:
\begin{equation}
\label{purinv3k}
\fl
f_{[\sigma_1,\dots,\sigma_{k-1}]_\approx}(\psi)=
\;\Tr
\prod_{\tau=ts^2,ts,t}
\Bigl(
\Id_{\{j\mid \sigma_j=\tau\}} 
\otimes
\Tr_{\{k\}\cup\{j\mid \sigma_j\in\{\tau,e\}\}}\pi^{T_{\{j\mid \sigma_j=s^2\}}}
\Bigr),
\end{equation}
where the $\prod$ product symbol means non-commutative product,
in the order of its subscript.
This gives back the formulae for the special cases $k=1$, $2$ and $3$.

\section{Mixed-state invariants}
\label{sec:mixinv}

In section~\ref{sec:luinvs}, we considered the mixed state invariants of $k$ subsystems
as pure state invariants of $k+1$ subsystems.
By considering the graphs of the invariants, given in section~\ref{sec:graphsops},
we can clarify this from another point of wiev.

An invariant is given by $\sigma\in S_m^k$:
this encodes an index-contraction for the matrices of the operators
$\pi=\cket{\psi}\bra{\psi}$ or $\varrho$ for pure or mixed states, respectively.
If $\sigma \not\approx \sigma'$ for $\sigma,\sigma'\in S_m^k$,
then they give rise to different polynomials for mixed states,
while it can happen, that $\sigma \sim  \sigma'$,
so they give rise to the same polynomial for pure states.
In this case, the unlabelled graphs given by $[\sigma]_\approx$ and $[\sigma']_\approx\neq[\sigma]_\approx$
are related to each other
by the independent permutation of the heads and tails of the edges,
while the corresponding operation is
the independent permutation of the coefficients $\psi$ and $\overline{\psi}$ in (\ref{purinv0}).
This operation is not allowed for mixed states.
In section~\ref{sec:pureinv},
we have given the decompositions of $\sim$-equivalence classes into $\approx$-equivalence classes
for some grade $m$ and for some $k$ numbers of subsystems,
leading to the different writings of the same polynomial.
For mixed states of $k$ subsystems, these polynomials are not the same anymore.
This offers us a different point of wiev, which seems to be more natural:
let us consider the pure state invariants as the special cases of the mixed state invariants
instead of considering the mixed state invariants as pure state invariants of a bigger system.
We have the mixed state formula (\ref{mixinv}) for the set of invariants, 
if we substitute a pure state $\cket{\psi}\bra{\psi}$ into them, then some of them will coincide,
but we can keep this in hand.

For the sake of completeness, we show the mixed state polynomials (\ref{mixinv}) below.
Comparing these formulae with the ones for pure states,
one can see 
how the $k$-partite mixed state invariants are related to the $k+1$-partite pure state ones 
(\ref{purmix}),
or,
how the different $k$-partite mixed state invariants coincide with the same $k$-partite pure state invariants.

Let $\varrho\equiv\varrho_{12\dots k}$ be a density matrix on $\mathcal{H}$.
The polynomials are labelled by $[\sigma]_\approx\in S_m^k/S_m$.

\subsection{Invariant polynomials of grade $m=1$ (degree one)}
For $m=1$, we have for all $k$
\begin{equation}
\label{mixinv1k}
f_{[e,\dots,e]_\approx}(\varrho) = \Tr\varrho_{12\dots k}.
\end{equation}

\subsection{Invariant polynomials of grade $m=2$ (degree two)}
For $m=2$, 
for one-partite system, ($k=1$, $\varrho\equiv\varrho_1$,) we have
\begin{eqnarray*}
 f_{[e]_\approx}(\varrho) &= (\Tr\varrho_1)^2,\\
 f_{[t]_\approx}(\varrho) &= \Tr\varrho_{1}^2,
\end{eqnarray*}
(for a $n_1=2$ one-qubit system, the determinant is an element of this subspace:
$2\det \varrho = f_{[e]_\approx}(\varrho)-f_{[t]_\approx}(\varrho)$)
for two-partite system, ($k=2$, $\varrho\equiv\varrho_{12}$,) we have
\begin{eqnarray*}
 f_{[e,e]_\approx}(\varrho) &= (\Tr\varrho_{12})^2,\\
 f_{[e,t]_\approx}(\varrho) &= \Tr\varrho_2^2,\\
 f_{[t,e]_\approx}(\varrho) &= \Tr\varrho_1^2,\\
 f_{[t,t]_\approx}(\varrho) &= \Tr\varrho_{12}^2,
\end{eqnarray*}
for three-partite system, ($k=3$, $\varrho\equiv\varrho_{123}$,) we have
\begin{eqnarray*}
 f_{[e,e,e]_\approx}(\varrho) &= (\Tr\varrho_{123})^2,\\
 f_{[e,e,t]_\approx}(\varrho) &= \Tr\varrho_3^2,\\
 f_{[e,t,e]_\approx}(\varrho) &= \Tr\varrho_2^2,\\
 f_{[e,t,t]_\approx}(\varrho) &= \Tr\varrho_{23}^2,\\
 f_{[t,e,e]_\approx}(\varrho) &= \Tr\varrho_1^2,\\
 f_{[t,e,t]_\approx}(\varrho) &= \Tr\varrho_{13}^2,\\
 f_{[t,t,e]_\approx}(\varrho) &= \Tr\varrho_{12}^2,\\
 f_{[t,t,t]_\approx}(\varrho) &= \Tr\varrho_{123}^2,
\end{eqnarray*}
and for arbitrary number of subsystems, ($\varrho\equiv\varrho_{12\dots k}$) we have
\begin{equation}
\label{mixinv2k}
 f_{[\sigma_1,\dots,\sigma_k]_\approx}(\varrho)
=\Tr(\Tr_{\{j\mid \sigma_j=e\}}\varrho)^2.
\end{equation}
The number of these---the dimension of the grade $m=2$ subspace of the inverse limit of algebras---is $2^k$.
The set of algebraically independent generators
contains all the $m=2$ polynomials from (\ref{mixinv2k}),
except the ones for which there are only $e$'s in $[\sigma_1,\dots,\sigma_k]_\approx$ labelling the polynomial.
The number of these is $2^k-1$.

\subsection{Invariant polynomials of grade $m=3$ (degree three)}
For $m=3$,
for one-partite system, ($k=1$, $\varrho\equiv\varrho_1$,) we have
\begin{eqnarray*}
 f_{[e]_\approx}(\varrho) &= (\Tr\varrho_1)^3,\\
 f_{[t]_\approx}(\varrho) &= \Tr\varrho_1 \Tr\varrho_1^2,\\
 f_{[s]_\approx}(\varrho) &= \Tr\varrho_1^3,
\end{eqnarray*}
(for a $n_1=3$ one-qutrit system, the determinant is an element of this subspace:
$6\det \varrho = f_{[e]_\approx}(\varrho)-3f_{[t]_\approx}(\varrho)+2f_{[s]_\approx}(\varrho)$)
for two-partite system, ($k=2$, $\varrho\equiv\varrho_{12}$,) we have
\begin{eqnarray*}
 f_{[e,e]_\approx}(\varrho) &= (\Tr\varrho_{12})^3,\\
 f_{[e,t]_\approx}(\varrho) &= \Tr\varrho_{12} \Tr\varrho_2^2,\\
 f_{[t,e]_\approx}(\varrho) &= \Tr\varrho_{12} \Tr\varrho_1^2,\\
 f_{[t,t]_\approx}(\varrho) &= \Tr\varrho_{12} \Tr\varrho_{12}^2,\\
 f_{[e,s]_\approx}(\varrho) &= \Tr\varrho_2^3,\\
 f_{[s,e]_\approx}(\varrho) &= \Tr\varrho_1^3,\\
 f_{[s,s]_\approx}(\varrho) &= \Tr\varrho_{12}^3,\\
 f_{[t,s]_\approx}(\varrho) &= \Tr(\Id_1\otimes\varrho_2)\varrho_{12}^2,\\
 f_{[s,t]_\approx}(\varrho) &= \Tr(\varrho_1\otimes\Id_2)\varrho_{12}^2,\\
 f_{[t,ts]_\approx}(\varrho)&= \Tr(\varrho_1\otimes\varrho_2)\varrho_{12},\\
 f_{[s,s^2]_\approx}(\varrho) &= \Tr(\varrho_{12}^{T_1})^3\equiv \Tr(\varrho_{12}^{T_2})^3,
\end{eqnarray*}
and for arbitrary number of subsystems ($\varrho\equiv\varrho_{12\dots k}$) we have
\begin{equation}
\label{mixinv3k}
f_{[\sigma_1,\dots,\sigma_k]_\approx}(\varrho)=
\;\Tr
\prod_{\tau=ts^2,ts,t}
\Bigl(
\Id_{\{j\mid \sigma_j=\tau\}}  \otimes
\Tr_{\{j\mid \sigma_j\in\{\tau,e\}\}}\varrho^{T_{\{j\mid \sigma_j=s^2\}}}
\Bigr),
\end{equation}
where the $\prod$ product symbol means non-commutative product,
in the order of its subscript.
This gives back the formulae for the special cases $k=1$ and $2$.

\section{Algorithm for $S_3^r/S_3$}
\label{sec:alg}
The formula in (\ref{purinv3k}) gives grade $m=3$ invariant polynomials for 
a $(\sigma_1,\dots,\sigma_{k-1}) \in S_3^{k-1}$ $k-1$-tuple of permutations,
but the linearly independent ones are labelled by $[\sigma_1,\dots,\sigma_{k-1}]_\approx \in S_3^{k-1}/S_3$.
Since the group-structure of $S_3$ is not too complicated,
we can give an algorithm
to construct exactly one representative element 
$\sigma$ for all orbits $[\sigma]_\approx$,
i.e., to construct the elements of $S_3^r/S_3$.
The choice $r=k-1$ and $r=k$ gives the labels for pure and mixed state invariants, respectively.

Again, $S_3=\{e,s,s^2,t,ts,ts^2\}$, $s=(123)$, $t=(12)(3)$,
and its conjugacy classes are $[e]=\{e\}$, $[s]=\{s,s^2\}$, $[t]=\{t,ts,ts^2\}$.
First, we write the conjugation table: $\beta,\gamma\in S_3$
\begin{equation*}
\beta\gamma\beta^{-1}:\qquad
\begin{array}{l||l|ll|lll}
\beta,\gamma & e   & s   & s^2 & t & ts & ts^2\\
\hline
\hline
e   & e  & s   & s^2 & t   & ts  & ts^2\\
s   & e  & s   & s^2 & ts  & ts^2& t   \\
s^2 & e  & s   & s^2 & ts^2& t   & ts  \\
t   & e  & s^2 & s   & t   & ts^2& ts  \\
ts  & e  & s^2 & s   & ts^2& ts  & t   \\
ts^2& e  & s^2 & s   & ts  & t   & ts^2\\
\end{array}
\end{equation*}

For every position of the list $(\sigma_1,\dots,\sigma_r) \in S_3^r$
there is a conjugacy class $[\sigma_j]$ of $S_3$, which remains unchanged under simultaneous conjugation.
For a given $(\sigma_1,\dots,\sigma_r)$, we would like to single out one representative element
in the orbit of simultaneous conjugation.
To do this, we examine the orbits.
\begin{itemize}
\item If $\sigma_j=e$ for all $j$, we have the trivial orbit of length $1$.
Besides this case, we do not have to deal with the positions in which $e$'s occur,
since $\sigma_j=e$ remains unchanged under simultaneous conjugation.
\item If $\sigma_j\in[s]$ (besides $e$) for all $j$, 
then we can choose the element of $[\sigma]_\approx$ which has $s$ in the first position in which an element of $[s]$ occurs.
These orbits are of length $2$.
\item If $\sigma_j\in[t]$ (besides $e$) for all $j$, then we have two kinds of orbits.
If $\sigma_j$ is the same for all $j$ for which $\sigma_j\in[t]$,
then we can choose the element which has $t$ in the first position in which an element of $[t]$ occurs.
These orbits are of length $3$.
On the other hand, 
if there are at least two different $\sigma_j$ for which $\sigma_j\in[t]$,
then it is not enough to fix only one position.
It can be checked by the conjugation table above
that for the ordered pairs of different elements of $[t]$
there exists \emph{exactly one} permutation which brings them into $(t,ts)$ by simultaneous conjugation.
So we can uniquely choose the elements which have
$t$ in the first position in which an element of $[t]$ occurs and
$ts$ in the first position in which a different element of $[t]$ occurs.
These orbits are of length $6$.
\item If both $[s]$ and $[t]$ occur (besides $e$), 
then we have to fix two positions again.
It can be checked by the conjugation table above
that for every pair given by the elements of $[s]\times[t]$
there exists \emph{exactly one} permutation which brings it into $(s,t)$ by simultaneous conjugation.
So we can uniquely choose the element which has
$s$ in the first position in which an element of $[s]$ occurs and
$t$ in the first position in which an element of $[t]$ occurs.
These orbits are of length $6$.
\end{itemize}

With the help of the observations above, we can formulate the following algorithm
generating $S_3^r/S_3$, i.e., the labels of the polynomials.
\begin{enumerate}
\item For every position of the list $(\sigma_1,\dots,\sigma_r) \in S_3^r$,
assign one of the conjugacy-classes of $S_3$.
Do this in all possible ways, and apply the following steps for all of them.
\item Write $e$ into all positions to which $[e]$ has been assigned.
\item Take the first of the positions to which $[s]$ has been assigned, and write $s$ there. 
To the others of such positions, write either $s$ or $s^2$ in all possible ways.
\item If there is no position with $[s]$, then
take the first of the positions to which $[t]$ has been assigned, and write $t$ there.
To the following of such positions, write either $t$ or $ts$ in all possible ways,
but after the occurrence of the first $ts$, write either $t$, $ts$ or $ts^2$ in all possible ways.
On the other hand,
if there is at least one position with $[s]$, then
take the first of the positions to which $[t]$ has been assigned, and write $t$ there. 
To the others of such positions, write either $t$, $ts$ or $ts^2$ in all possible ways.
\end{enumerate}

What is the number of the labels obtained in this way?
This could be find by the use of some combinatorics, but we do not have to follow that way.
If the local dimensions $3\leq n_j$, 
then the elements of $S_3^{k-1}/S_3$ label the 
linearly independent grade $m=3$ invariants,
and their number, the dimension of the grade $m=3$ subspace of the inverse limit of the algebras
is given in~\cite{HW,Peti23}.
For $m=3$ pure state invariants, this is
$|S_3^{k-1}/S_3| = 6^{k-2}+3^{k-2}+2^{k-2}$,
for mixed states
$|S_3^k/S_3| = 6^{k-1}+3^{k-1}+2^{k-1}$
(see also in~\cite{oeis}).
One can easily check that 
the set of algebraically independent generators
contains all the $m=3$ polynomials from (\ref{purinv3k}) or (\ref{mixinv3k}),
except the ones for which---using the labelling algorithm above---%
there are only $e$'s and $t$'s in $[\sigma]_\approx$ labelling the polynomial.
(This is the way for the permutations not to act transitively on the set of $m=3$ labels.)
The number of these is 
$6^{k-2}+3^{k-2}+2^{k-2}-2^{k-1}=6^{k-2}+3^{k-2}-2^{k-2}$ for pure states, and
$6^{k-1}+3^{k-1}-2^{k-1}$ for mixed states.

\section{Summary and notes}
\label{sec:summary}
In this paper we have written out explicitly
the LU-invariant polynomials
for pure and mixed states,
given in (\ref{purinv}) and (\ref{mixinv}), for $m=1,2,3$.
This was done for arbitrary number of subsystems of arbitrary dimensions.
The key point---and the new feature---here is the independency \cite{HWW,Peti23}: 
the polynomials in (\ref{purinv}) and (\ref{mixinv})
give the \emph{linearly independent} basis of the $m$ graded subspace of the algebra of LU-invariant polynomials
if $m\leq n_j$ (for all $j$) \cite{HWW},
and some of them---the ones for which the defining permutations together act transitively---%
become an \emph{algebraically independent} generating set 
in the inverse limit of algebras, i.e., if $n_j\to\infty$ for all $j$ \cite{Peti23}.
This independency result shows the power of the elegant approach using the inverse limit of the algebras of LU-invariant polynomials.

However, for a given $n=(n_1,\dots,n_k)$ system,
it seems to be usual~\cite{LPpureorbs,LPSmixedorbs}
that it is not enough to use only the polynomials of maximal degree $2m$, 
where $m\leq n_j$ for all $j$, for the separation of the LU-orbits.
(According to the relatively simple case of $n=(2,2,2)$ three qubits,
where it is known~\cite{Sudbery3qb,Acin3qbPureCanon},
that we need an $m=3$, an $m=4$, and an $m=6$ invariant polynomial---%
the Kempe invariant, the three-tangle, and the Grassl-invariant respectively,---beyond the $m\leq2$ ones.)
If $m\nleq n_j$ for a $j$,
the generators given in (\ref{purinv}) and (\ref{mixinv}) will not be linearly independent,
and the algebraic relations between them exhibit a complicated structure.

New results are given by the nice compact formulae of grade $m=3$ invariants for pure and mixed quantum states,
(equations (\ref{purinv3k}) and (\ref{mixinv3k}), respectively)
and in the algorithm generating the different equivalence classes of permutation $k$-tuples of $S_3$
under simultaneous conjugation, given in section~\ref{sec:alg}.
The latter is necessary to eliminate identical polynomials.
Connections between pure and mixed state invariant polynomials has been illustrated as well.
These results are obtained by the use of graphs corresponding to the polynomials \cite{HWW,Peti23}.

Note, that
(i) the same degree of the pure state invariants in the coefficients and in their complex conjugate,
(ii) the much simpler labelling of the mixed state invariants than the pure ones,
(iii) considering the pure state invariants as the special cases of the mixed ones,
seem to stress that 
the mixed states (density matrices) are the natural objects in the topic of unitary invariants
instead of the pure states (state vectors).
This approach is widely supported by the whole quantum physics,
where the elements of the lattice of the subspaces of the Hilbert space
are regarded to be more fundamental than the elements of the Hilbert space themselves.

\begin{figure}\centering
\includegraphics{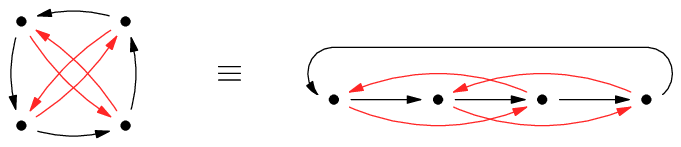}
\caption{An example for an $m=4$, $k=2$ mixed state LU-invariant polynomial 
which can not be written by the considered simple matrix operations.}\label{fig:imp}
\end{figure}

The illustrating polynomials given in this paper
could have been written in a convenient form using 
partial trace, matrix product, tensorial product and partial transpose
for grade $m=1$, $2$ and $3$.
However, we note that
it can happen that a grade $m\geq4$ invariant polynomial
can not be written by using these operations only.
At this time, we can not formulate general necessary and sufficient conditions for this,
but we can give an enlightening example.
For the use of matrix operations, we have to write down the matrices one after the other,
this fixes the order of the vertices in some sense.
The partial traces form loops of edges.
If we can find an ordering of the vertices (up to cyclical permutations),
in which these loops of every colours contain only adjacent points (with respect to this ordering)
then the matrix-operations can be written for the entire polynomial.
This situation can be seen in the third row of figure~\ref{fig:mxopgraphs}.
After some drawing, one can check that 
there is no such an ordering of the vertices for the graph in figure~\ref{fig:imp},
which seems to be the most simple exapmle for such a situation.

\ack
We thank P\'eter Vrana and P\'eter L\'evay for helpful discussions,
and Markus Grassl for informations.
This work was partially supported by the New Hungary Development Plan
(Project ID: T\'AMOP-4.2.1/B-09/1/KMR-2010-0002).

\section*{References}

\bibliography{cubicinvs4}{}
\bibliographystyle{unsrt.bst}

\end{document}